\documentstyle[preprint,prb,aps]{revtex}

\begin{document}

\draft
\title {Distribution Function Analysis of Mesoscopic Hopping 
Conductance Fluctuations}

\author{R. J. F. Hughes$^{\dagger \S}$, 
A. K. Savchenko$^{\ddagger}$, 
J. E. F. Frost$^{\dagger}$, E. H. Linfield$^{\dagger}$, 
J. T. Nicholls$^{\dagger}$,
M.~Pepper$^{\dagger}$, E. Kogan$^{\S}$ and M. Kaveh$^{\S}$ }

\address{$\dagger$ Cavendish Laboratory, Madingley Road, Cambridge CB3 
0HE, U.K.\\
$\S$ Minerva Center, Jack and Pearl Resnick Institute 
of Advanced Technology, Dept. of Physics, Bar-Ilan University, 
Ramat-Gan 52900, Israel.\\
$\ddagger$ Department of Physics, University of Exeter, Stocker 
Road, Exeter EX4 4QL, U.K.}

\date{\today }

\maketitle
\begin{abstract}

Variable-range hopping (VRH) conductance fluctuations in the 
gate-voltage characteristics of mesoscopic GaAs and Si
transistors  are analyzed by means of their full distribution
functions (DFs).  The forms of the DF predicted by the theory of
Raikh and Ruzin  have been verified under controlled conditions
for both the long,  narrow wire and the short, wide channel
geometries. The variation  of the mean square fluctuation size
with temperature in wires  fabricated from both materials is
found to be described quantitatively  by Lee's model of VRH
along a 1D chain. Armed with this quantitative validation  
of the VRH model, the DF method is applied to the  problem  of
magnetoconductance in the insulating regime. Here a
non-monotonic variation of the
magnetoconductance is observed  in Si MOSFETS whose sign at
low magnetic fields is dependent  on the channel geometry. 
The origin of this defect is discussed within the framework of
the interference model of VRH magnetoconductance in terms of 
narrowing of the DF in a magnetic field.
\end{abstract}

\pacs{PACS numbers: 72.20.-i,72.20.My,73.40Qv,73.50.Jt}

\section{Introduction}

Mesoscopic conductance fluctuations in the insulating regime  of
small, disordered transistors were first observed by Pepper 
\cite{pep} in GaAs MESFETs and then studied in detail in Si
MOSFETs  by Fowler, Webb and coworkers \cite{fow} in the early
1980s. Extremely  strong random fluctuations, spanning several
orders of magnitude,  were observed at low temperatures in the
conductances of narrow-channel  devices as the gate voltage was
varied. At first the origin of  this effect was not clear. Azbel
\cite{azb} suggested that resonant  tunneling from source to
drain might produce such structure as  the chemical potential
was swept through transmission resonances  of the eigenstates in
the random impurity potential. However,  it became clear that
this zero-temperature mechanism could not  make a significant
contribution to the conductance in the relatively  long devices
of this experiment. Instead, the most satisfactory  explanation
was provided by Lee \cite{lee} who proposed a model in which 
electrons move by variable-range hopping (VRH) along a
one-dimensional  (1D) chain. A number of elementary hopping
resistances, each  depending exponentially on the separation and
energy difference  between sites, are added in series to give
the overall resistance  of the chain. In this model it is
assumed that, because of the  extremely broad distribution of
the elementary resistors, the  total chain resistance can be
well approximated by that of the  single most resistive hop. The
fluctuations then arise as a consequence  of switching between
the pairs of localized sites responsible  for the critical hop
as each elementary resistance reacts differently  to a change in
the chemical potential. These fluctuations are  therefore of
``geometrical'' origin, arising from the random  positioning of
localized sites in energy and space, as distinct  from the
``quantum'' nature of the tunneling mechanism which  would be
strongly affected for example by an applied magnetic  field.

Serota, Kalia and Lee \cite{ser} went on to simulate the
ensemble  distribution of the total chain resistance  $R$ and
its dependence on the temperature  $T$ and the sample length
$L$. In their ensemble, the random impurities are distributed 
uniformly in energy and position along the chain. In
experiments  a single device is generally used, so that the
impurity configuration  is fixed, and fluctuations are observed
as a function of some  variable external parameter such as the
chemical potential. An  ergodicity hypothesis is then invoked to
the effect that the  same ensemble is sampled in both cases,
something that has been  verified experimentally by Orlov  {\it
et al.} \cite{orl1}. Using the natural logarithm of the
resistance,  the authors of Ref. \onlinecite{ser} 
obtained for the mean and standard deviation:

\begin{equation}
\label{1}
\langle \ln~R \rangle \sim \left(\frac{T_0}{T}\right )^{1/2}
\left [\ln\left(\frac{2L}{\xi}\right)\right]^{1/2}
\end{equation}
\begin{equation}
\label{2}
s\equiv\langle \left(\ln~R- \langle \ln~R \rangle\right )^{2}
\rangle \sim \left(\frac{T_0}{T}\right )^{1/2}
\left [\ln\left(\frac{2L}{\xi}\right)\right]^{-1/2}
\end{equation}
where $\xi$ is the localization radius and $T_{0}$ is the
characteristic temperature for Mott VRH:  $T_{0}~=~1/k_{B} \rho
\xi$ ($\rho$ is the density of states at the Fermi energy). It
can  be seen that the size $s$ of the fluctuations decreases
extremely slowly with  length, a result characteristic of 1D
which was first pointed  out by Kurkijarvi \cite{kur}. The
explanation is simply that exceptionally  large resistance
elements, even though they may be statistically  rare, dominate
the overall resistance since they cannot be by-passed  in this
geometry. The averaging assumed in the derivation of  Mott's
hopping law for 1D does not occur and the total resistance 
takes on the activated form of the largest individual element.

A detailed analytical treatment of this model was undertaken  by
Raikh and Ruzin \cite{rai1,rai2} who divided the problem up into
a number  of length regimes. Their theory introduces the concept
of the  ``optimal break'', the type of gap between localized
states (on  an energy versus position plot) which is most likely
to determine  the overall resistance. The optimal shape of such
a state-free  region has maximal resistance for the smallest
area and turns  out to be a rhombus. A sufficiently long chain
will have many  such breaks in series to give a most probable
resistance

\begin{equation}
\label{3}
R_{\mbox{prob}}=R_0\frac{L}{\xi}
\left(\frac{T_0}{T}\right )^{1/2}\exp
\left(\frac{T_0}{2T}\right )^{1/2}
\end{equation}
where $R_{0}$ is the prefactor in the Mott VRH formula.  This
formula breaks down once the expected number of optimal  breaks
in the chain becomes small, of order one. It is valid  only when
$\nu \gg 1$, where $\nu$ is a parameter defined implicitly by

\begin{equation}
\label{4}
\nu= \frac{2T_0}{T}\ln \left(\frac{L\nu^{1/2}}{\xi}\right)
\end{equation}
For $\nu < 1$, which corresponds to the normal experimental 
situation, the resistance of the chain is determined by a few 
sub-optimal breaks of which the expected number occurring in  a
chain of length $L$ is approximately one. The most probable
resistance,  or its logarithm $Q$, then follows a more
complicated temperature law of the form:

\begin{equation}
\label{5}
Q\equiv\ln\frac{R}{R_0}=\frac{\nu^{1/2}T_0}{T}\approx
\left\{2\frac{T_0}{T}\ln\left[ \frac{L}{\xi}
\left(\frac{T_0}{T}\right )^{1/2}
\ln^{1/2}\left(\frac{L}{\xi}\right)\right]\right\}
\end{equation}

The probability distribution function (DF) for the quantity 
$f(Q)$ is best written in terms of $\nu$ and a new parameter
$\Delta$. For $\nu <1$ it is given by the following integral:

\begin{equation}
\label{6}
f\left(\Delta\right)=\frac{e^{\Delta}}{\pi}\int^{\infty}_0 dx
\exp\left(-x^{\nu^{1/2}}\cos\frac{\pi\nu^{1/2}}{2}\right)
\cos\left(xe^{\Delta}-x^{\nu^{1/2}}\sin\frac{\pi\nu^{1/2}}{2}
\right)
\end{equation}
\begin{equation}
\label{7}
\Delta \equiv Q-\frac{\nu^{1/2}T_0}{T}
\end{equation}
$f(\Delta)$ is a function with a peak close to $\Delta = 0$
 and width 
determined by $\nu$. In principle it can be written in terms of 
$\xi$ and $\rho$ using Equations (\ref{4}) through (\ref{7}). 
There is a simple relationship 
between $\nu$ and the variance of $Q$:

\begin{equation}
\label{8}
\langle Q^2 \rangle - \langle Q \rangle^2=\frac{\pi^2}{6}
\left(\frac{1}{\nu}-1\right)
\end{equation}
This theory is equally applicable \cite{rai3} to the case of the
transverse  conductance $G$ of a thin film or barrier. Instead
of a sum of series  resistances, the required quantity is the
sum of parallel conductances  representing conducting chains of
hops traversing the film. Whereas  in 1D the total resistance is
determined by the blocking effect  of the critical hop, here the
total conductance is dominated  by an optimal ``puncture'': an
uncommonly high-conductance hopping  chain through the barrier
which effectively shorts out all other  current paths. On a
logarithmic scale, since $\ln~R~=~-\ln~G$, the DFs for the two
geometries are simply reflections  of each other. The variation
of the width and peak position with  $\xi$ and $\rho$ is
different, however, in the two cases. Fig. 1 shows  schematics
of the two geometries, the equivalent resistor networks  and DFs
from Equation (\ref{6}) characteristic of each case. Here,  as
throughout the remainder of the paper, the abscissa is $\ln~G$.
In 1D the importance of blocking resistors adds weight  to the
contribution of extremely high resistances and produces  a long
tail out to low values of $\ln~G$. For a short 2D barrier the DF
has the opposite asymmetry  with a tail out to high
conductances, reflecting the effect of  punctures in shorting
out less conductive paths. In fact the  form of the DF is
universal, the theory requiring only that the  elementary
quantities to be summed are independent and come from  an
exponentially wide distribution. The microscopic details of  the
conduction mechanism enter only into the dependence of $\nu$ and
$\Delta$ on external parameters such as the temperature  and magnetic
field. The requirement of independence in the case  of the
barrier means that conductive chains must be sufficiently  far
apart, which should be satisfied for a barrier with a
sufficiently  large aspect ratio $W/L$. 
We use the description ``short 2D'' for this short-, 
wide-channel geometry to distinguish it from the square 2D geometry 
in which conduction is via an interconnected percolation network.

While the above theory is that best suited to the hopping
regime,  numerous authors \cite{abr} have examined
the problem of the {\it zero-temperature} conductance
distribution of disordered  wires and they find that, for a
sufficiently long wire, the DF  is Normal in $\ln~G$. Some
authors \cite{coh} argue that this is also the case  for the
insulating regime in higher dimensions. To obtain a
finite-temperature  result, it is necessary to introduce a
temperature-dependent  coherence length. Kramer  {\it et al}.
\cite{kra,mar} have used the Mott hopping length for this 
purpose and performed an average over localization lengths to 
calculate the size of the fluctuations in $\ln~G$. (The Lee
model does not allow for a distribution of  localization
lengths). With the caveat that it is difficult to  rigorously
justify the averaging method, they find that the size  of the
fluctuations varies with temperature in $d$ dimensions as

\begin{equation}
\label{9}
s\simeq\left(\frac{T_0}{T}\right)^{1/2(d+1)}
\end{equation}
For the case $d~=~1$, in which we are mostly interested, 
this yields 
$s~\sim~T^{-1/4}$, in contradiction with equation (2) which 
predicts $s~\sim~T^{-1/2}$. An intermediate model \cite{lad}
describes fluctuations that are partly geometrical, partly
quantum coherent in nature.

The experimental results presented below are able to 
distinguish between these two theories, providing  good
quantitative  agreement with the geometrical, 1D hopping chain
 theory. Section
III introduces  the results on 1D devices in both silicon and
gallium arsenide  and analyses the temperature dependence of the
conductance and  its fluctuations in terms of these theories.
Section IV exhibits  some experimental DFs which are also well
described by the theory,  both for the 1D and short 2D
geometries with their opposite characteristic  asymmetric
distributions. It is shown that fits to the full DF  are more
reliable in studying the fluctuation amplitude than  just the
standard deviation which is prone to large statistical  errors
due to the long tails. Finally, Section V examines the  effect
of an applied magnetic field and suggests interesting 
possibilities for investigating the complicated area of hopping 
magnetoconductivity by means of experimental DFs.

\section{ Devices and Experimental Method}

A number of different devices, both Si and GaAs and in both the 
1D and short 2D geometries, were used for this study. The Si 
MOSFETs were small-area CMOS devices with an oxide thickness  of
210A and self-aligned ohmic contacts. The 2DEG is  formed by
inversion in the very lightly-doped p-type Si adjacent  to the
oxide barrier. Lithographic dimensions for the 1D devices 
ranged from 0.5 to 2$\mu$m wide by 5 to 20$\mu$m long for the 1D
wires  and 100$\mu$m wide by 1 to 2$\mu$m long for the short 2D
geometry. Here  the designation of length $L$ always refers to
the distance between source and drain.  The electronically
active dimensions of the channel were estimated  from
measurements of samples of varying size to be about 0.5$\mu$m
smaller than these nominal values.

The GaAs devices were fabricated from a simple delta-doped 
layer of Si donors with a Hall carrier concentration of  $4
\times 10^{11}$ cm${}^{-2}$. Channels were defined by the
application  of a negative bias to patterned surface gates. The
split gate  method \cite{tho} was used to define 1D wires, while
narrow strip gates  were used to create short 2D barriers.
Beyond the channel definition  voltage, increasing the bias
serves to shift the chemical potential  in the active region of
the device without greatly changing the  channel dimensions.
Because of the spreading out of the electric  field between the
edges of the patterned gate and the delta-layer  situated
0.3$\mu$m
below the surface, the dimensions of the electrically  active
regions differ significantly from the lithographic dimensions. 
For example, a split gate with gap dimensions 1mm square
produces  a narrow channel approximately 0.2$\mu$m wide and
1.8$\mu$m long. This  estimate is based on the observation that
split gate devices  narrower than 0.8mm did not conduct. Strip
gates 30$\mu$m wide by  0.5 to 2$\mu$m long were used to produce
short 2D barriers in the  region of 0.8$\mu$m longer than the
lithographic thickness. Aside  from the ill-defined channel
dimensions, a further problem with  these devices was the high
series resistance resulting from the  partially depleted 2DEG
region near the gates. For the short  2D devices it was
estimated that this series resistance, which  varies with gate
voltage, would be sufficient to truncate some  of the
high-conductance fluctuations. Our measurements were taken 
after brief illumination with light from a red LED, using the 
persistent photoconductivity effect to reduce the series 
resistance from the sample leads.

Two-terminal a.c. resistance measurements were taken in
dilution  refrigerators down to temperatures of 50mK. For the Si
devices,  a low-frequency (8-18Hz) excitation voltage of 5 or
10$\mu$V was  used, depending on the temperature, and saturation of
temperature-dependent  quantities was observed below about 70mK.
The temperature dependence  of, for example, the average
conductance flattens out quite abruptly  below this temperature.
Measurements have been carried out using  several different
experimental set-ups and it is our experience  that the
saturation temperature increases monotically with the  observed
noise level. For the GaAs devices it was necessary to  use a
higher excitation voltage to obtain a sufficiently good 
signal-to-noise ratio and saturation of the average conductance 
and the fluctuation amplitude was evident below 200mK. Because 
of this and the above-mentioned problems most of the data
presented  here are from Si MOSFETs. However, all of the results
in sections  III and IV for Si have been reproduced in the GaAs
devices, with  the same quantitative agreement but somewhat
greater uncertainties.

\section{Temperature Dependence of the Mean Conductance and 
Fluctuation Amplitude}

The first part of the experiment is to determine the Mott
parameter  $T_{0}$ from an analysis of the temperature
dependence  of the mean value of the conductance. This is for
comparison  with the value of $T_{0}$ needed to explain the size
of the fluctuations.  Although we expect VRH to be the
conduction mechanism, in our  mesoscopic device the averaging
implicit in Mott's law is not  taking place. We therefore first
take the logarithm of the measured  conductance and then perform
a numerical average of the data-points  over a suitable range of
gate voltage to to obtain the quantity $<\ln G >(=-<\ln R>)$.
 It is important  to reach a
suitable compromise between choosing a gate voltage  interval
sufficiently small that $T_{0}$ does not vary significantly over
its length  yet sufficiently wide to meet the requirements of
statistical  accuracy. More will be said on this matter in
section III on  DFs where the choice is much more critical.
However, thanks to  the large number of fluctuations observed at
low temperatures,  these requirements are easily met.

Experimental data for a long, narrow 19.4 by 0.6$\mu$m Si
MOSFET  channel are presented in Fig. 2, where a section of the
gate  voltage characteristic has been split into seven intervals
for  averaging. It is not possible to distinguish between
activation  laws with inverse temperature exponents of 1/3, 1/2
or even 1  from the temperature range we have available.
However, there  does appear to be a change of gradient in the
vicinity of 0.3K.  This could be indicative of a change from 1D
to 2D Mott hopping  as the temperature is raised and the hopping
distance becomes  shorter than the sample width. Indeed, if the
values of $T_{0}$ above 0.3K are extracted using the 2D formula 
with exponent 1/3, and values below 0.3K using the 1D formula 
with exponent 1/2, equating the hopping lengths in the two
formulae  yields a crossover temperature of order 0.5K, roughly
consistent  with the observed crossover point. If we introduce
the known  sample width $W$, this interpretation allows $\xi$
and $\rho$ to be obtained independently. These work out at 
$\xi~\sim {}~0.1\mu$m and $\rho~ \sim {}~0.3$ times the density
of states for a 2D subband  in silicon. Both these values are
physically reasonable and consistent  with the conduction
mechanism being VRH and 1D in nature below  0.3K. We therefore
make this assumption and restrict ourselves  to the sub-0.3K
range in our subsequent DF analysis. Note that  the large value
of $\xi$ shows that we are not very deep in the insulating
regime.

The next task is to analyze the temperature dependence of the 
fluctuation amplitude. We take the same gate voltage intervals 
as before and this time calculate the standard deviation $s$ of
$\ln~G$. Here a log-log plot of $s$ against $T$ does not yield a
straight line (hollow circles, Fig.  5(c)). The slope varies
from approximately $-0.5$ to $-1.5$ depending  on the temperature.
This upper power is much greater than can  be accounted for by
any of the theories. However, the standard  deviation as
calculated directly from the data points is dominated  by the
contribution from the low-conductance tail of the distribution, 
where noise and statistical uncertainty is greatest. As a
result,  we postpone further comment until the next section
where a full  distribution function analysis can resolve the
issue. The problem  is not so apparent in a gallium arsenide
split-gate device estimated  as being 0.2$\mu$m wide by 
1.8$\mu$m long.
Here the data points (of which  there are only five ranging from
0.2 to 0.5K) best fit the law  $s~ \sim {}~0.4 T^{-0.6 \pm
0.1}$. This supports the model of Lee  and of Raikh and Ruzin
expressed in equation (2) at the expense  of the scaling
approach in equation (9) which would predict a  power of $-1/4$.
Moreover, estimating $\xi~ \sim ~0.1\mu$m from the measured
value of  $T_{0}~=~1.6$K and the assumption that the value  of
$\rho$ is roughly the 2D subband density of states multiplied 
by the channel width, gives the length-dependent prefactor in 
equation (2) as  $\left [\ln \left (2 L/\xi \right )\right
]^{-1/2}~=~0.5$,  very close to the observed prefactor  of 0.4.
The exact numerical coefficient in Equation (\ref{2}) is not 
given by the authors, but our data indicates that it should be 
close to unity. Thus the limited available gallium arsenide
data  appears to be perfectly describe by the VRH chain model.
For  silicon, once the overall shape of the distribution is
taken  into account, both the power and the numerical
coefficient are  found to be perfectly in line with the
calculations \cite{ser,rai1} based  on the 1D VRH model. This
will be the conclusion of the following  section.

\section
{Experimental Distribution Functions: Geometry and Temperature 
Dependence}

An analysis relying solely on the first two moments of the
distribution  of $\ln~G$ gives only limited information and,
where the distribution  has long tails as in this case, may give
unrepresentative or  erroneous results. The aim of this paper is
to show that analysis  of the full DF from experimental
fluctuations in mesoscopic devices  can be an important tool for
investigating hopping in the mesoscopic  regime and may shed
light on the processes underlying the macroscopic 
magneto-conductivity. To start with, we describe our method  of
obtaining an experimental DF histogram.

The raw data consists of a set of points representing the
fluctuations  in conductance as a function of some external
parameter, usually  a gate voltage $V_{g}$. It is important to
establish that there  is no zero offset error since the first
step is to take the logarithm  of the conductance. Generating
the desired histogram is simply  a matter of binning the data
points into suitable classes of  $V_{g}$, whose width is chosen
as a compromise  between acceptable resolution and statistical
error. The statistical  uncertainty is governed not by the total
number of data points  but by the number of conductance
fluctuations in the data-set.  If the number of points covering
each peak in the characteristic  is excessively high, the data
may be thinned out before binning  by retaining only every 
$n$th point. Care needs to be taken in identifying a ``noise 
floor'' in the conductance measurement, below which point the 
data is meaningless. Such unreliable data cannot be included  in
the histograms since taking the logarithm magnifies the noise 
at low conductances and might produce an artificially long tail 
in $\ln~G$. Aside from these considerations, the critical
decision  is in the gate voltage range of the data-set used to
generate  the histogram. This cannot be made arbitrarily wide
since it  is the nature of the devices that $\rho$ and $\xi$
vary slowly with $V_{g}$ as the chemical potential sweeps
through  the localized states at the edge of the band (whether
it be conduction  or impurity band in nature). If this variation
causes the mean  or background conductance level to change
significantly over  the chosen interval of $V_{g}$, the DF will
be smeared out. This was a  problem for Orlov  {\it et
al}.\cite{orl2} who studied the DF of short 2D channels down  to
1.2K in GaAs devices similar to our own. To obtain a sufficient 
number of fluctuations they had to subtract a smooth background 
from $\ln~G$, something that cannot be strictly justified and
which  does fully solve the problem. Our gate voltage
characteristics,  measured at much lower temperatures, have the
advantage of much  denser fluctuations and it is possible to
obtain histograms of  a satisfactory quality without any
background subtraction. As  a rough indication, a $V_{g}$ range
covering 15 or more peaks yields  a good histogram provided that
the background conductance does  not vary by more than about
10-20\% of the total distribution  width.

An example of DFs calculated from the experimental gate-voltage 
characteristic of a 0.2$\mu$m wide by 1.8$\mu$m long GaAs device
is reproduced  in Fig. 3 from Ref. \onlinecite{hug1}. The
characteristic has been split into  five intervals so that the
variation of the DF as the channel  is pinched off can be seen.
The decreasing conductance and increasing  relative size of the
fluctuations translates simply into a distribution  which shifts
to lower $\ln~G$ and broadens so that it is fit by a theoretical
DF from  equation (6) with a smaller value of $\nu$. The
peak position shifts quite markedly between adjacent  intervals,
suggesting that the choice of interval size is close  to its
maximum reasonable limit. Unusually, this data shows a 
non-monotonic variation of the peak position with gate voltage. 
The effect is specific to this device and its cause is unclear. 
A final important point is that the experimental limitations  on
the smallest conductances which can be measured above the  noise
level severely restricts the usable gate voltage range.  Thus a
significant portion of the last DF in the figure is already 
lost in the noise. At the other end of the range, the
conductance  does not have to get very high before the
fluctuations weaken  and the assumption of an exponentially
broad distribution breaks  down: it is for this reason that the
first DF does not have a  significant 1D tail. Characteristics
for DF analysis therefore  have to lie within this gate voltage
window, implying that it  is only possible to study a narrow
range of values of  the parameter  $T_{0}$.

To fit the experimental DFs to equation (6), the integral is 
calculated on a mainframe computer 
for all values of $\nu$ with a resolution of 0.01.
 These results are stored 
and used to perform a two-parameter least-squares fit to the 
experimental histogram by shifting the computed curve along the 
$\ln~G$ axis until the optimum fit to the peak position 
$\Delta_{0}$ (which is more precisely the difference 
between $\Delta$ and $ln~G$) is found. This is repeated for each value of %
{\it n} to find the best global least squares fit to $\nu$
 and $\Delta_{0}$.

The results confirm the Raikh and Ruzin theory for the form of 
the DF under controlled 1D conditions. Other authors have
previously  examined the DF in the short 2D geometry, notably
Orlov  {\it et al}. \cite{orl2} in n-type GaAs and Popovich 
{\it et al}.\cite{pop} who claim to have observed a weak
asymmetry  in a large-area, wide Si MOSFET. Yakimov  {\it et
al}. \cite{yak} observed a crossover from the 1D to short  2D
forms for conduction across thin a-Si films. The thinnest  films
apparently showed 1D behavior, which is surprising given  that
here the 2D DF theory should be most applicable. Interestingly, 
these authors obtain their histograms without relying on the 
ergodicity hypothesis by measuring a large number of
macroscopically  identical devices. In this paper we find
quantitatively good  fits to both distributions under controlled
conditions using  differently shaped devices on the same chip.

Figure 4 shows three histograms obtained from three MOSFETs
fabricated  on the same chip. The middle graph shows a
reasonable 1D DF,  though the situation is not ideal owing to
the rather large channel  width and almost metallic
conductances. The topmost graph shows  good agreement between
the short 2D theory and experimental data  from a device 2$\mu$m
long by 100$\mu$m wide. However, good agreement  for this short
2D case is not as ubiquitous in our data as it  is for the 1D
wires. Frequently the DFs show no clear asymmetric  tail at high
conductances and are well represented by a Gaussian  curve, as
in the lower graph. The reasons for this have not been  resolved
but some observations may provide hints as to possible 
explanations. The simplest explanation would be to claim that 
the devices possessed macroscopic inhomogeneities which
produced  fixed high-current paths and therefore destroyed the
desired  wide-channeled geometry. However, it is not true that
some devices  produce good 2D DFs and others are somehow
imperfect; it is rather  that the 2D asymmetric distribution is
only seen in certain regimes.  In particular, the asymmetric
distribution is observed only,  in the most insulating
measurable gate voltage range when the  distribution is very
wide. The narrow distributions occurring  in the more weakly
insulating part of the characteristic are  invariably symmetric
although the converse is not true: sometimes  the most resistive
part of the characteristic still yields symmetric 
distributions. Inexplicably, a strong magnetic field of the
order  of 5-10T helps to restore the asymmetric distribution in
these  cases. This has been observed in both silicon and GaAs
devices.  It is perhaps possible that there are other mechanisms
at work  in the tail of the distribution where the conductance
is generally  rather high to be truly in the VRH regime. Some
such mechanism  - perhaps related to resonant tunneling or the
increased importance  of electron correlations at these higher
currents - could be  responsible for truncating the high
conductance tail in many  of our measurements. This question is
still unresolved, although  we can be certain that the effect is
not due to the addition  of any series contact resistance since
subtracting even the largest  plausible values for such a
resistance before calculating $\ln~ G$ has little effect on the
shape of the distributions.

In order to reach the final result of this section, we
re-examine  the temperature dependence of the fluctuation size
using a distribution  function analysis rather than simply
calculating the standard  deviation $s$ of the data points.
Comparison between the two methods  shows that DF fits obtained
by calculating $s$ and then using Equation (\ref{8}) to select
the appropriate  value of $\nu$ gives very poor results. This
method appears to give  good fits to the tail of the
distribution, where the experimental  histograms are prone to
large statistical errors, at the expense  of the bulk of the
distribution. Reanalysis of the 19.4$\mu$m long  Si MOSFET shows
that the discrepancy between the two methods,  while small for
the mean log conductance, can be as much as a  factor of two in
the standard deviation. Fig. 5(a) shows the  full set of DFs and
their fits to the 1D theory. On the accompanying  plots 5(b) and
(c), hollow circles represent simple calculations  of the mean
and standard deviation directly from the data points  while full
circles are the equivalent quantities calculated from  the best
fit to the entire DF. Either method results in a similar 
estimate of the Mott parameter of $T_{0}~=~6-8$K (Fig. 5(b)).
For the standard deviation  $s$, the straightforward method as
we saw above gives a  $T^{-1.4}$ power law dependence saturating
at low  temperatures (in GaAs at higher temperatures the best
power law  was $T^{-0.6}$). The full DF analysis, however,
yields  a straight line on a log-log plot over the entire range
with  a best fitting power also of $-0.6$ (Fig. 5(c)). This is
very close  to the prediction of $-0.5$ for the 1D chain model and
once again  disagrees with the result derived from scaling. In
Fig. 5(d)  the fit to a forced -0.5 power law gives a prefactor
of 0.35,  very close to the value of 0.4 from the values of
$\rho$ and $\xi$ estimated in section III. So, in both Si and
GaAs devices,  Lee's 1D chain hopping chain model gives a
quantitatively accurate  description of the results, but a
reliable analysis requires  the distribution function method

\section{Variation of the DF with magnetic field}

Having verified the applicability of both the 1D hopping chain 
theory and Raikh and Ruzin's form of the DF to our devices, it 
is now possible to use the experimental histograms as a tool 
with which to investigate the mechanisms of magneto-conductance 
in the VRH regime. This field has witnessed a large theoretical
effort, particularly using the interference mechanism of Nguen,
Spivak and Shklovskii \cite{ngu} (see review \cite {shk} and also
Refs.\onlinecite{med3,zha,sch,rai4} for more recent calculations). In this model the
presence of scatterers between the initial and final sites
allows interference to occur between alternative possible paths
within a single hop. The phase shifts introduced when a
perpendicular magnetic field is applied can give rise to a
positive magnetoconductance (PMC) in a macroscopic 2D or 3D
sample. The macroscopic conductivity is derived from a
logarithmic average over the distribution of elementary hopping
conductances, which becomes narrower in a magnetic field. At low
fields the peak of this distribution does not shift and the
relatively small PMC is a result of the extra weight afforded to
low conductances by the logarithmic average \cite{ngu}. At slightly
higher fields it is possible for the peak of the distribution to
shift to higher conductances, corresponding to a significant
increase in the localization length, and the PMC can be
exponentially large \cite{zha}.

Experiments \cite{lai,tre,ye,jia} on 2D gallium arsenide samples widely exhibit PMC in the VRH
regime, as does the study on indium oxide films \cite{ova}. The interference effects
are generally smaller in silicon devices but PMC has been observed in the 2D
inversion layer of MOSFETs \cite{har} and also in narrow-channelled, quasi-1D MOSFET
channels \cite{oha}. Our devices are not macroscopic so that the precise details of
the averaging procedure needed to calculate the magnetoconductance in a 2D
percolation network do not concern us. Instead, analysis of the mesoscopic
conductance fluctuations in terms of their distribution functions can give direct
insight into the elementary distributions at the root of the interference theory,
provided that the effect of sample geometry is taken into account. We have
already seen that the geometry effect in our samples is well-described by the
theory of Raikh and Ruzin. The present section therefore proceeds by introducing
some experimental magnetoconductance data, mostly from silicon MOSFET devices,
together with a suggestion as to its interpretation in terms of the elementary
hopping distribution.

As before, the first approach is to calculate the mean value 
$\langle \ln~G\rangle$ from a suitable section of the gate voltage characteristic. 
The variation of $\langle \ln~G\rangle$ with magnetic field $B$
 is obtained from data in which numerous gate voltage 
sweeps are taken at a series of fixed magnetic fields. Over the 
full magnetic field range, the gross behavior is similar in GaAs 
and Si both for 1D and short 2D devices. Fig. 6 shows the variation 
of $\langle \ln~G\rangle$ with $B$ in Si MOSFETs of both geometries up to high fields. 
At temperatures of order 100mK there is a NMC in fields up to 
about 1T by a factor of order 
{\it e} followed by a larger PMC up to fields of around 7T. 
The NMC is not observed in the GaAs devices but the same PMC 
is evident and is undoubtedly related to the proximity to the 
insulator-quantum Hall transition also observed in macroscopic 
devices made from these material \cite{hug2}. At higher fields, the 
conductivity rapidly freezes out. These exponentially large changes 
in conductance are the result of changes in the localization 
length induced by the magnetic field. At lower fields, of the 
order of a few tenths of a Tesla, a smaller magnetoconductance 
is observed in the silicon devices where this field range has 
been probed in detail. The effect is interesting in that the 
sign of the magnetoconductance is positive in the case of a 
1D wire and negative in the case of a short 2D sample. The presence 
or absence of such an effect has not been sufficiently investigated 
in the GaAs devices. Figs. 7 and 8 show the behavior of 
$\langle \ln~G\rangle$ in the two instances, together with DFs and samples 
of the data. In general, the time period necessary to collect 
all the data for a large number of magnetic fields is too great 
for the gate voltage sweeps to remain entirely reproducible -- 
this was the case in Fig. 6 and all our experiments to high fields. 
However, an acceptable degree of reproducibility was achieved 
for the data presented in the low-field figures. The detailed 
structure of the characteristic changes substantially over this 
field range. The overall relative magneto-conductance is in 
both cases by a factor of order 
{\it e}. Measurable conductances are only obtainable with hopping 
lengths of the order 0.1$\mu$m so, in this low-field range, 
the hopping 
length $r_{M}$ and the magnetic length are comparable. 
With the observed values of $T_{0}$ generally around 5K and a measurement temperature 
of just under 0.1K the hopping length should be several times 
greater than the localization length so that the condition 
$r_{M} \gg \xi$ demanded by most of the theories is met.
For the 1D device, our estimates suggest that $r_{M}$ is only
slightly greater than the channel width while $\xi$ is somewhat
smaller. We therefore believe that the hopping path is
effectively one-dimensional while the channel is wide enough to
accomodate the cigar-shaped coherence region of the interference
model.

It is not possible to demonstrate conclusively the origin of the different signs
observed in the low-field magnetoconductance for these two samples but the
following explanation appears the most attractive. According to the interference
model, the distribution of elementary hopping conductances narrows on applying a
small magnetic field. Our experimental DFs do not directly measure this
elementary distribution which is skewed because of the high aspect ratios of our
channel geometries. In 1D, because the overall conductance tends to be dominated
by difficult hops, the low conductance tail in the elementary distribution is
amplified in the experimental DF to produce the characteristic tail out to low
conductances. When the elementary distribution narrows slightly under the
influence of a magnetic field, the effect on the experimental DF should also be
amplified so that the long tail disappears very quickly. Conversely, for the
short 2D geometry, the experimental DF is biased in favor of high conductances so
that when the elementary distribution narrows, the high conductance tail in our
observed DF should rapidly disappear. The result is that a narrowing of the
elementary distribution in a magnetic field, provided that it is not accompanied
by a shift in the mean conductance, should result in a positive
magnetoconductance for a 1D wire but a negative magnetoconductance for a short
2D channel. This is exactly what we observe in the magnetic field dependence of
the average $<\ln·G>$ up to about 0.5T. However, a look at the distributions
themselves can confirm or disprove this hypothesis.

Taking the 1D case first, the experimental histograms in Fig. 7
immediately show that our hypothesis is reasonable. The
low-conductance  tail rapidly disappears to the point where the
fit to the theoretical  1D DF starts to become dubious. On the
other hand, the high-conductance  threshold of the distribution
does not shift at all, especially  when looking at the fitted
functions. The overall effect is a  shift in the weight of the
distribution leading to the PMC shown  in the adjacent plot (b).
The narrowing of the distribution is  apparent in the shrinking
of the standard deviation with B witnessed  in Fig. 7(c). Moving
next to the 2D case in Fig. 8, we see that  the real situation
is perhaps more complex. While it is arguable  that the high
conductance tail starts to disappear at about 0.5T,  the low
conductance threshold steadily decreases. In fact Fig.  8(c)
shows a slight  {\it broadening} of the distribution, completely
at odds with  the distribution-narrowing hypothesis and with the
theoretical  calculations. It should be noted that the
distributions are somewhat  noisy and the fits consequently
poorer so that it is difficult  to be quantitatively certain of
the results. It should also be  borne in mind that we do not
have a good explanation for the  NMC observed at fields of out
to 2T in both transistors. Both  MOSFETs show this, irrespective
of the channel geometry, despite  the fact that their
construction is not identical.

Evidently then, more questions are raised than are  answered. At
present the experimental data is very limited and  further
experiments should help to elucidate the magneto\-conductance 
mechanisms. Given the tantalizingly complicated non-monotonic 
variation of $\langle \ln G \rangle$ with $B$, further work is
clearly desirable. Extension of the  measurements to the
high-field regime will be difficult because  the conductance
changes by several orders of magnitude and also  because of
problems with reproducibility. However, the task could  be
approached piecewise by analyzing different sections of the 
gate voltage characteristic for different field ranges.
Detailed  examination of the low-field regime in GaAs devices is
also a  priority.

\section{Conclusion}

This paper details an experimental investigation into the use 
of distribution functions to analyze hopping conductivity in 
mesoscopic Si and GaAs transistors. A reliable method for obtaining 
histograms of useful quality has been described and the form 
of the DF predicted by Raikh and Ruzin for the 1D and short 2D 
cases verified under controlled conditions. For 1D Si wires, 
DF analysis is instrumental in showing that the temperature dependence 
of the fluctuation size is in agreement with the predictions 
of Lee's model for VRH along a 1D chain. The same quantitative
agreement is also found in  
GaAs wires. Finally, we have shown that experimental distribution
 functions from 
mesoscopic devices can provide a useful method of gaining 
additional 
insight into the mechanisms of magnetoconductance. 
In Si MOSFETs at low fields, the dependence of the sign of the 
magnetoconductance depends on the device geometry,
being positive in a narrow quasi-1D device and negative in a
short 2D channel. This is what one would expect from a narrowing
of the elementary hopping conductance  distribution taking into
account the selection effects of the channel geometry.
For the 1D device, 
DF analysis shows that the observed PMC in $<\ln G>$ is
explicable as a consequence of this narrowing of the elementary
distribution as predicted by the interference model. In the
slightly more complicated 2D case, this explanation is not
suppported by the experimental DFs and further experiments are
warranted.
\vskip 1cm

The authors are very grateful to G. Citver and V. Ginodman for 
technical assistance. We would like to thank D. H. Cobden, I. 
Shlimak and M. E. Raikh for illuminating discussions. RJFH acknowledges 
support from Trinity College Cambridge and the kind hospitality 
of Bar-Ilan University. We are grateful to Y. Oowaki for providing 
a Si MOSFET sample.

\begin{figure}
\caption{Device schematics and theoretical DFs for two values of the 
parameter n. (a) The 2D device behaves like a sum of parallel 
conductors and its DF has a tail out to high $\ln~G$. 
(b) The 1D device behaves like a sum of series resistances 
with a tail on the DF towards low $\ln~G$.}
\end{figure}

\begin{figure}
\caption{(a) Part of the experimental characteristic on a
logarithmic conductance scale from a $19.4\times 0.6\mu$m 1D 
Si MOSFET. Dotted lines 
demarcate gate voltage intervals used for averaging. (b) Fits 
of this data to the 1D Mott law averaged over the marked gate 
voltage intervals together with the values of $T_{0}$
 extracted below 0.3K.}
\end{figure}

\begin{figure}
\caption{ Conductance fluctuations from a $1.8\times 0.2\mu$m
1D GaAs 
device and experimental DFs obtained from five adjacent gate 
voltage intervals spanning the characteristic. At low gate voltages, 
the distribution is no longer exponentially wide while at high 
gate voltages the conductance becomes too small to measure. In 
between is a region where good fits to the theoretical 1D DF 
(solid curves) can be obtained.}
\end{figure}

\begin{figure}
\caption{ Distribution functions obtained from three Si MOSFETs 
fabricated on the same chip showing the characteristic 1D and 
2D asymmetries. Lithographic channel dimensions are (length  
width): (a) $2\times 100\mu$m fit by 2D DF, (b) $5\times 2\mu$m 
fit by 1D DF and 
(c) $1.5\times 100\mu$m fit by a Gaussian. Inset are the regions of the 
characteristic from which the histograms were obtained.}
\end{figure}

\begin{figure}
\caption{ Temperature dependence of the fluctuations from a 
$19.4\times  0.6\mu$m Si MOSFET. (a) Experimental DFs fit by
 the 1D theoretical 
form. (b) Fit to Mott's law of the average of $\ln~G$
 obtained both directly (hollow circles) and from the 
position $\Delta_{0}$ of the fitted DFs (filled circles). (c) Temperature 
dependence of the fluctuation amplitude $s$ both measured by the standard deviation of the data 
points (hollow circles) and calculated from the fits to the 1D 
DFs (filled circles). The gradient of the latter yields a 
$T^{-0.65}$ power law. (d) Fitting the fluctuation amplitude 
to a $T^{-1/2}$ power law yield the prefactor 0.35.}
\end{figure}

\begin{figure}
\caption{ Average magnetoconductance of Si MOSFETs with the (a) 
1D and (b) short 2D geometries. On this scale the low-field PMC 
is indicated by just the first two points in (a).}
\end{figure}

\begin{figure}
\caption{ (a) DFs obtained from a $19.4\times 0.6\mu$m Si MOSFET as a 
function of magnetic field and their fits to the 1D theory. (b) 
The average magnetoconductance obtained by the direct and fitting 
methods is positive. (c) The standard deviation of the distribution 
obtained by the two methods decreases.}
\end{figure}

\begin{figure}
\caption{ (a) DFs obtained from a $2\times 100\mu$m Si MOSFET as a function 
of magnetic field and their fits to the 2D theory. (b) The average 
magnetoconductance obtained by the direct and fitting methods 
is negative. (c) The standard deviation of the distribution obtained 
by the two methods increases slightly.}
\end{figure}

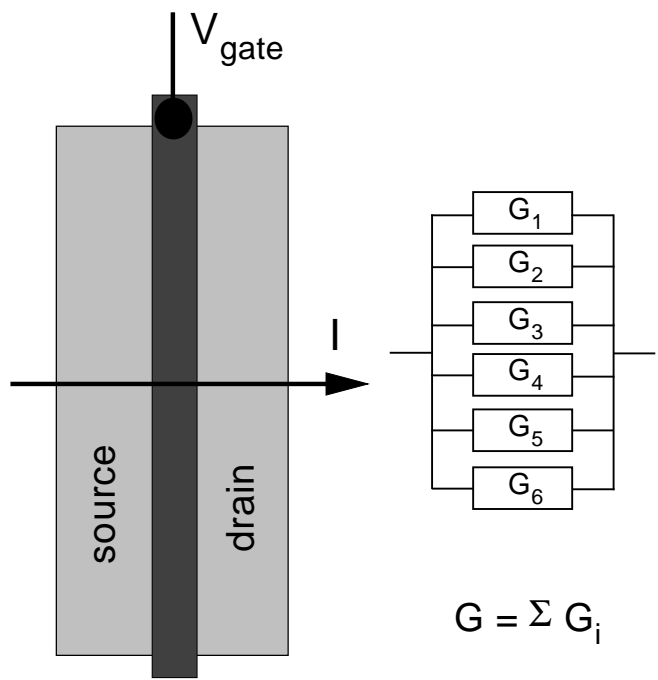
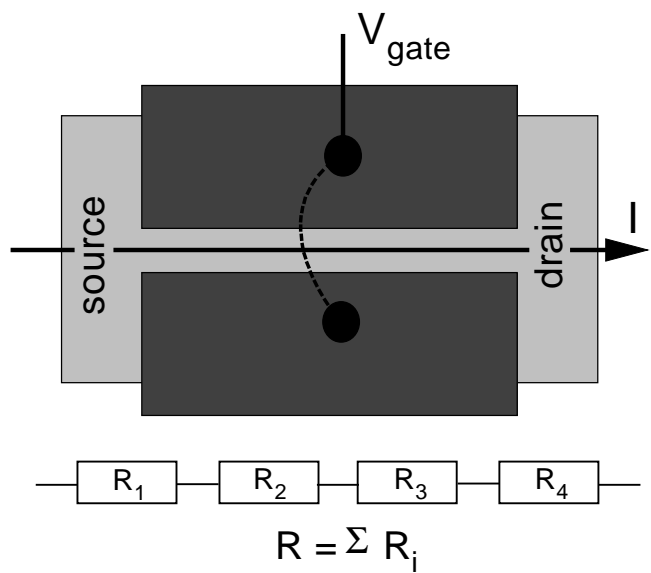
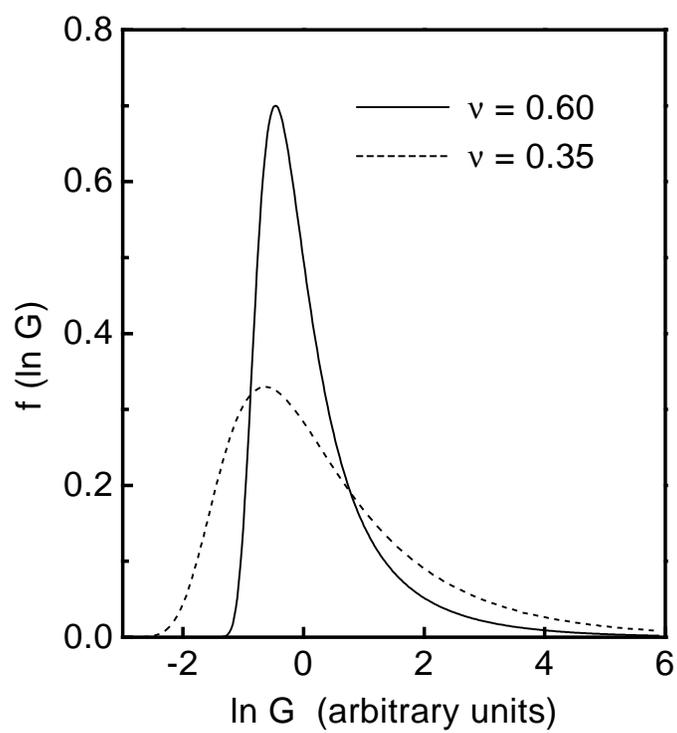
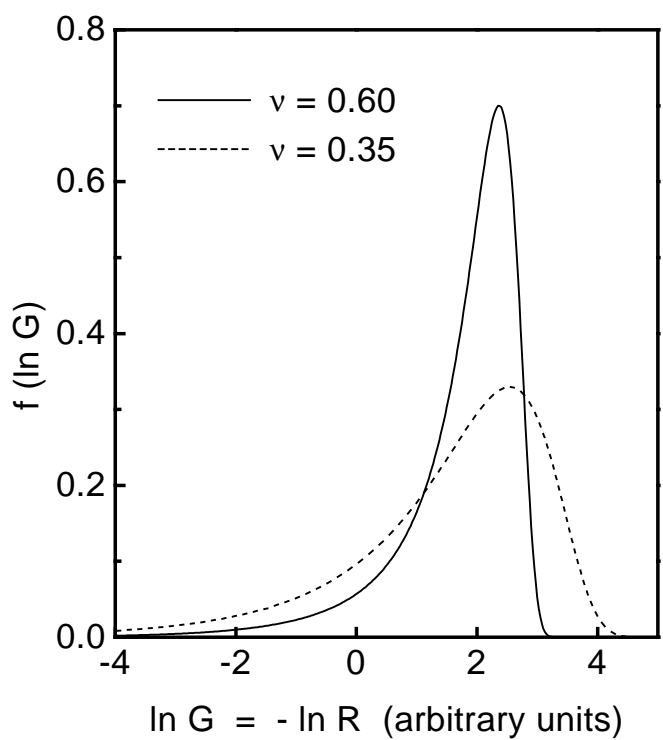
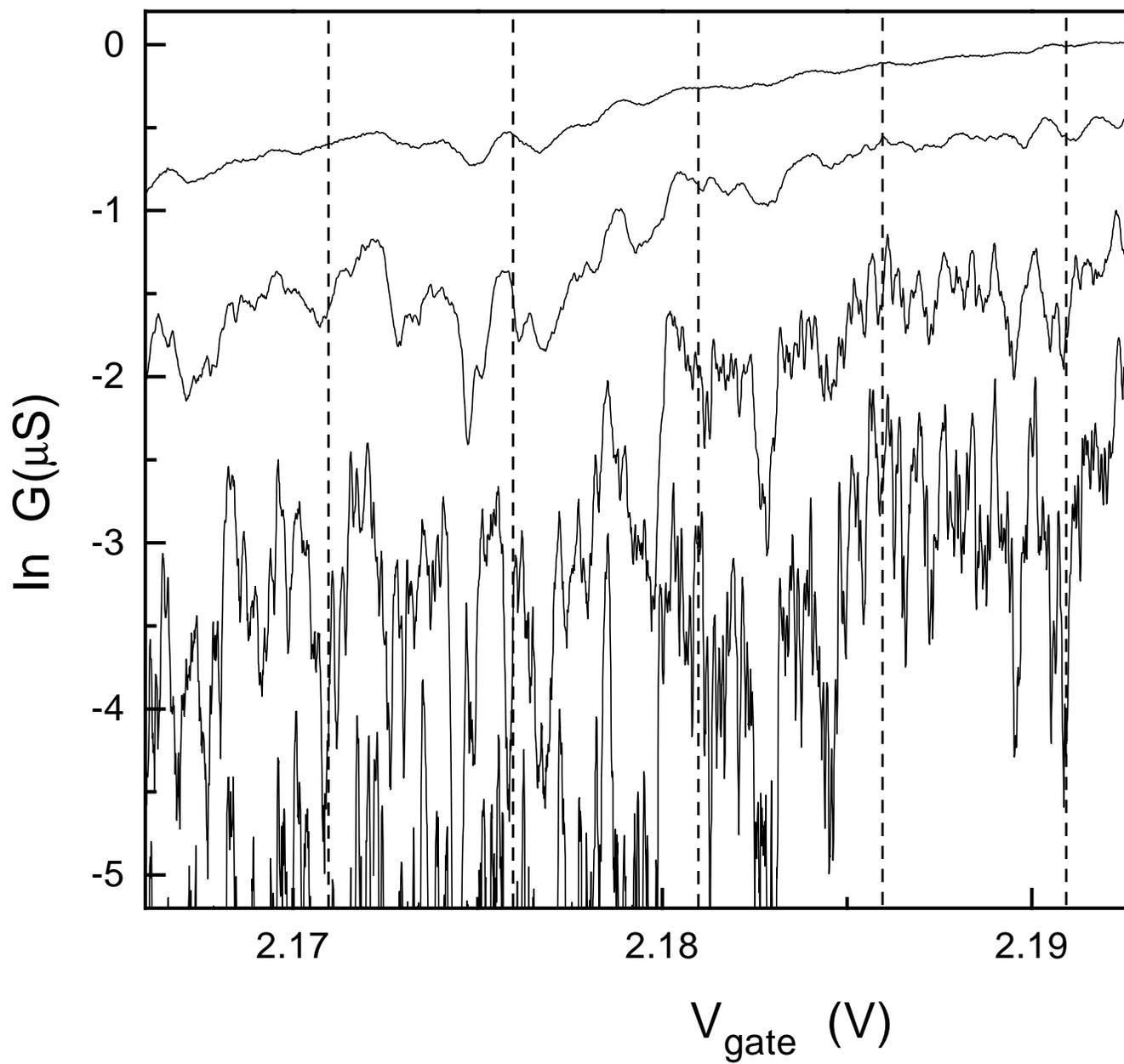
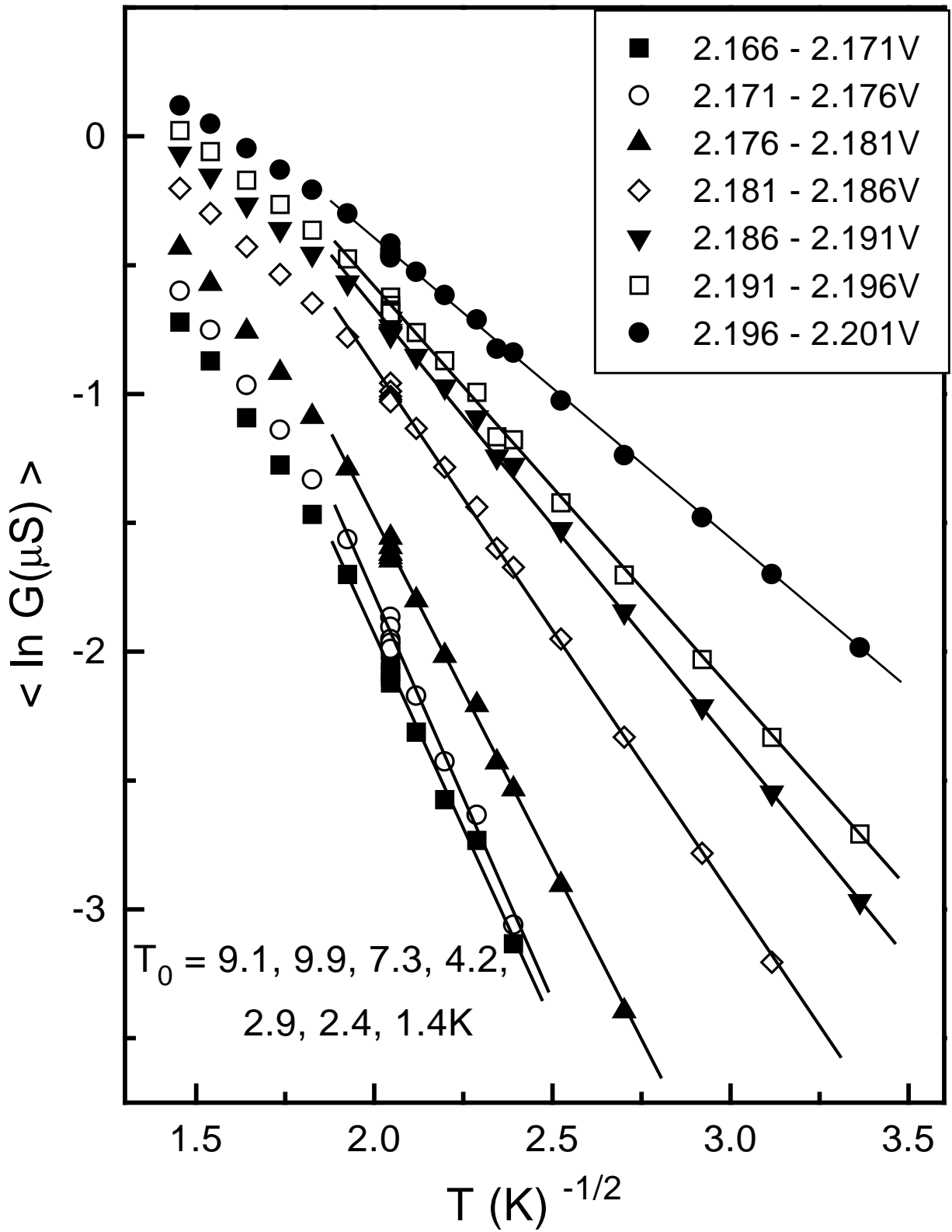
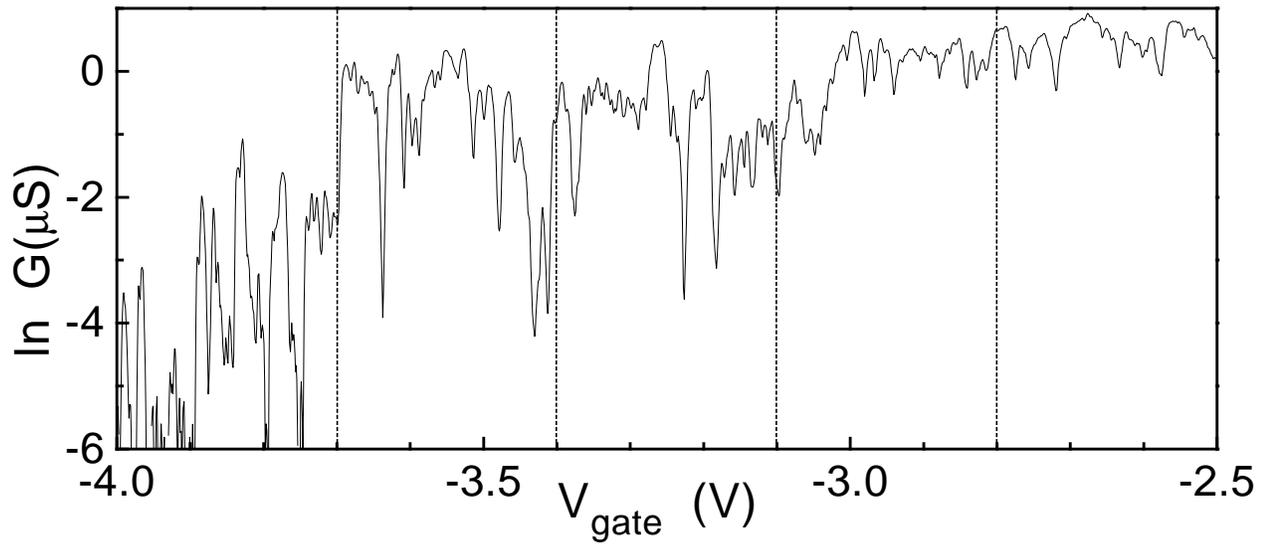
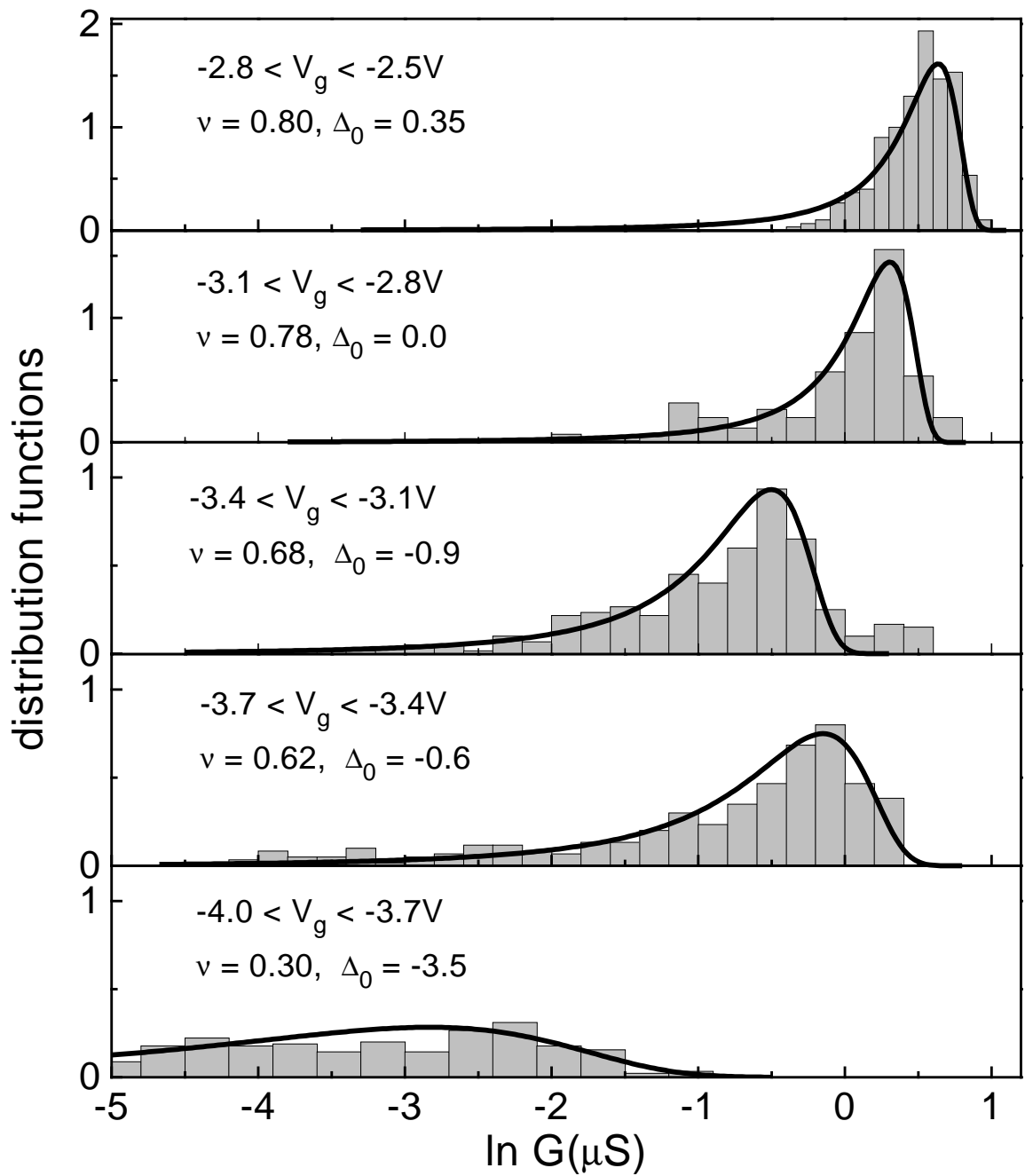
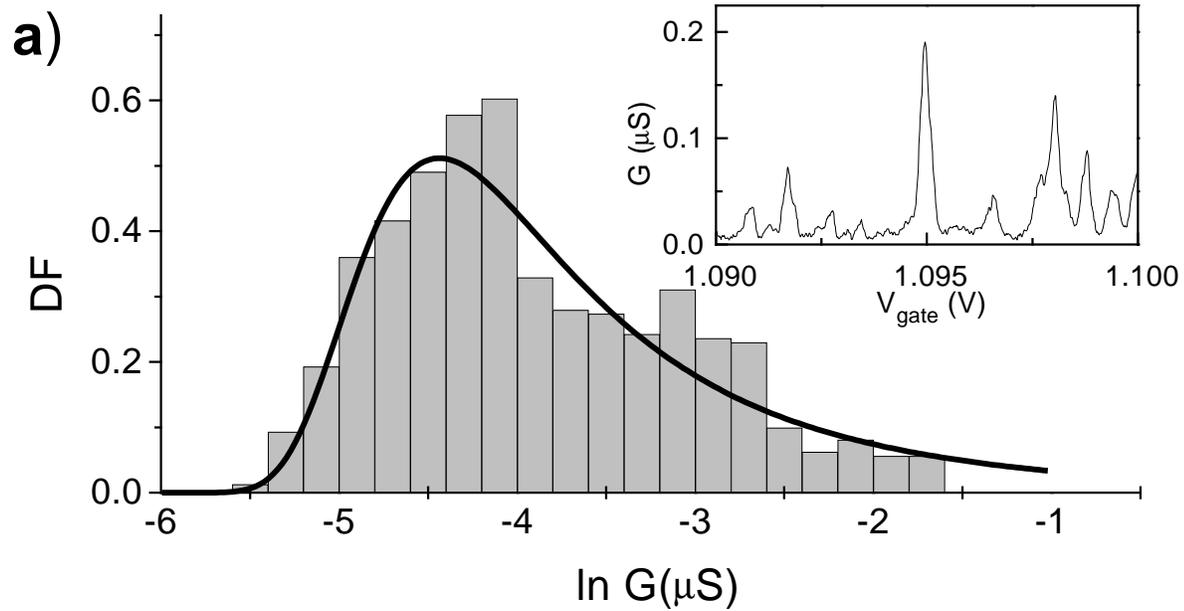
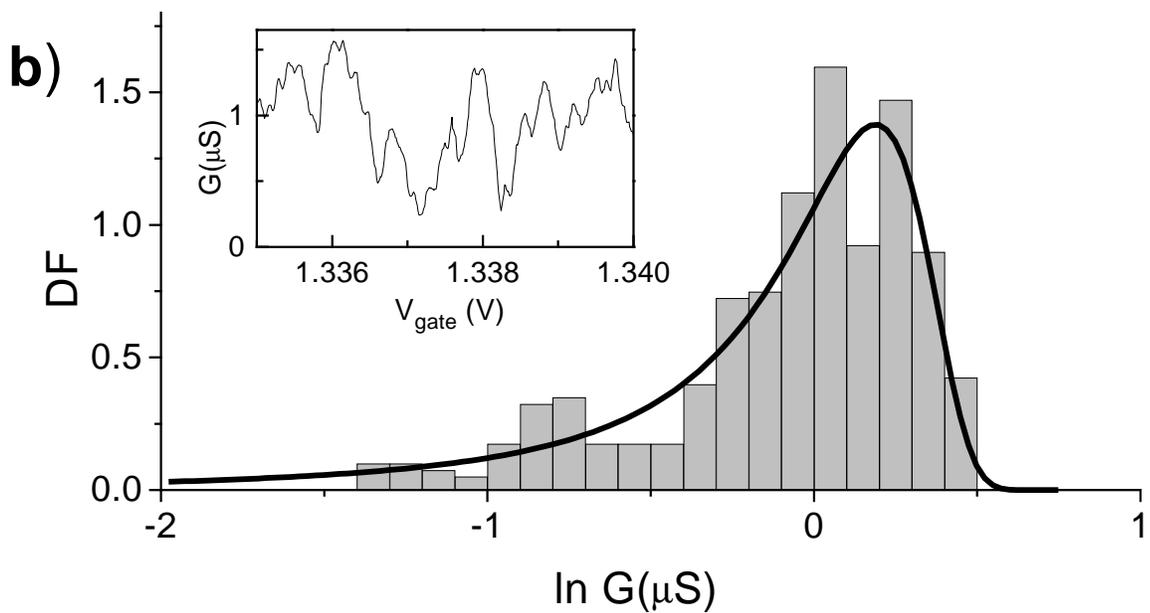
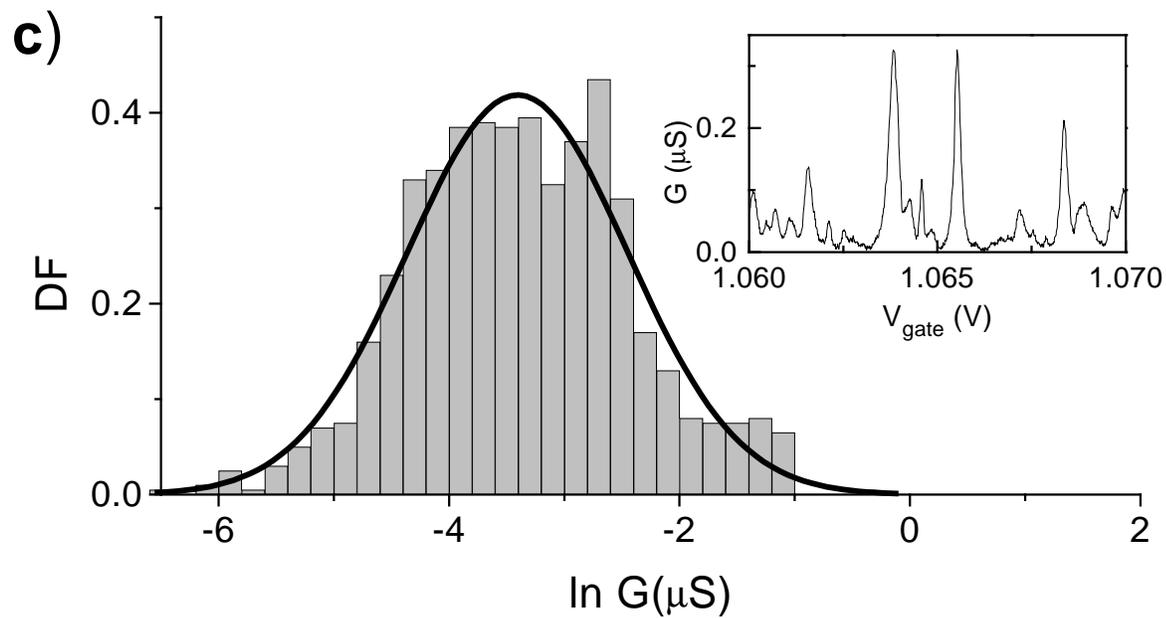
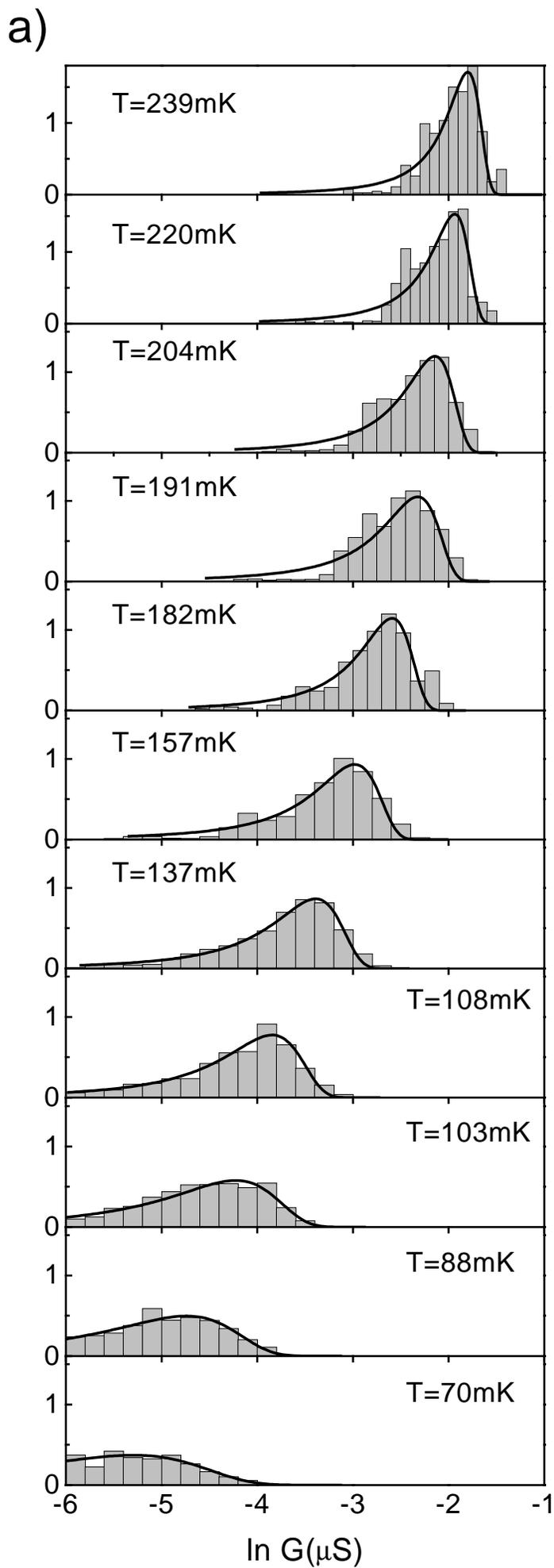
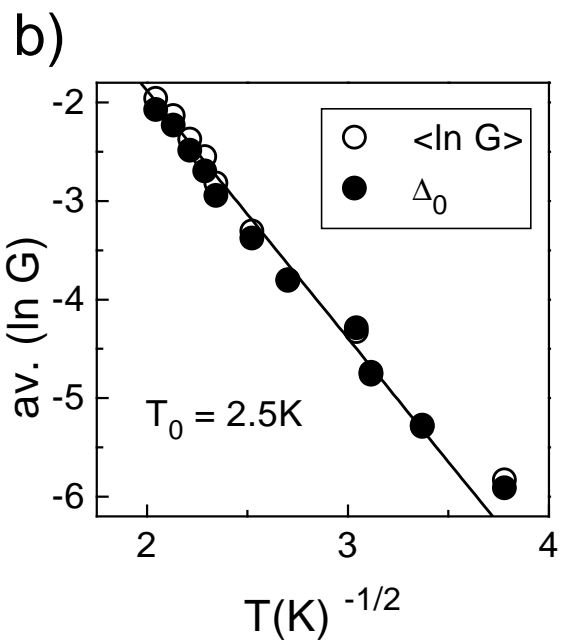
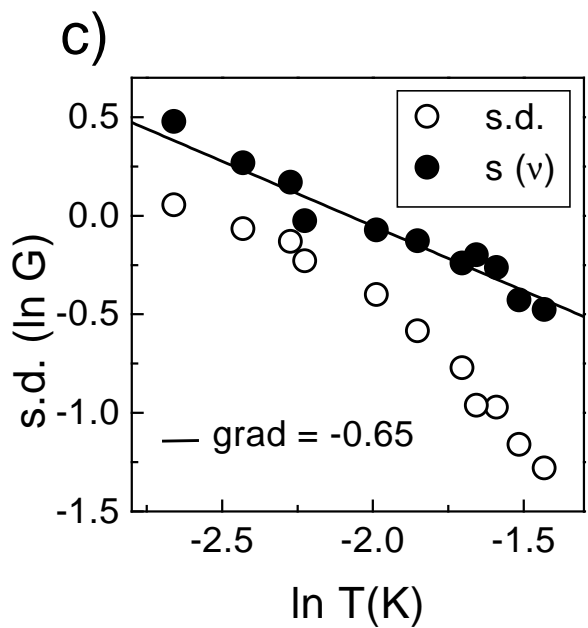
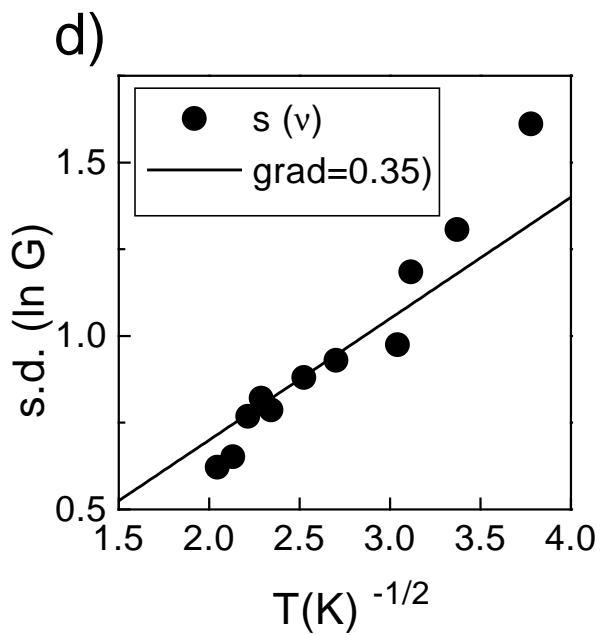
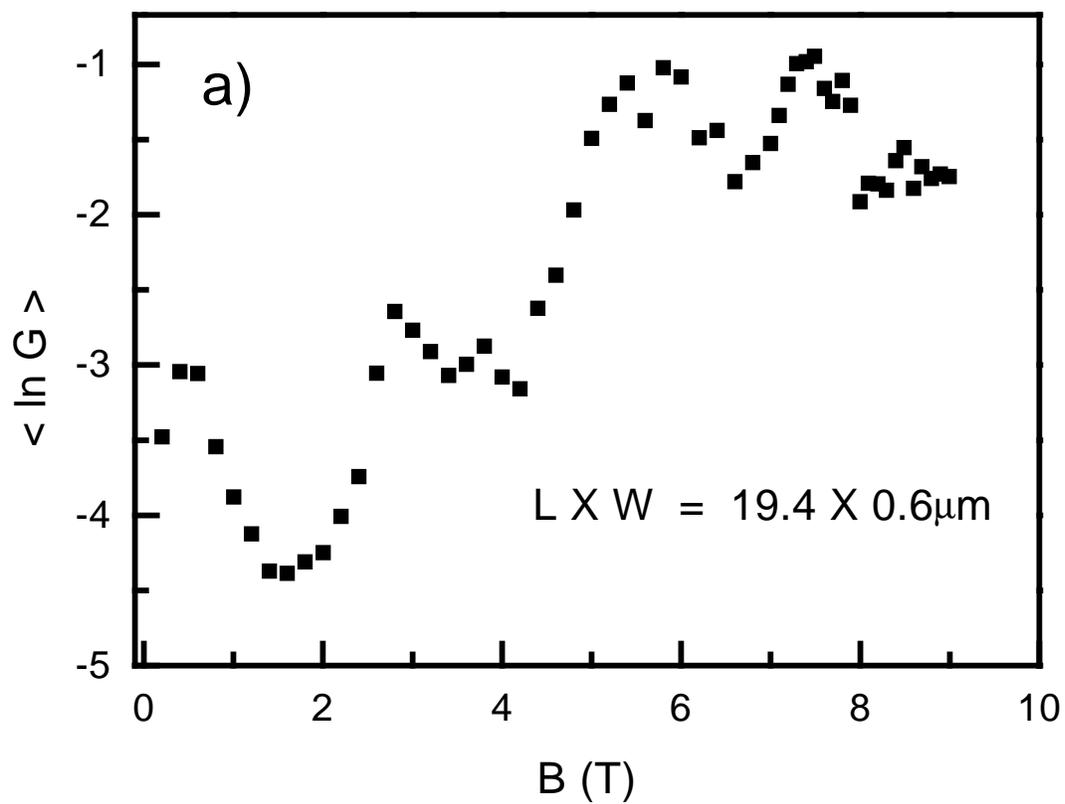
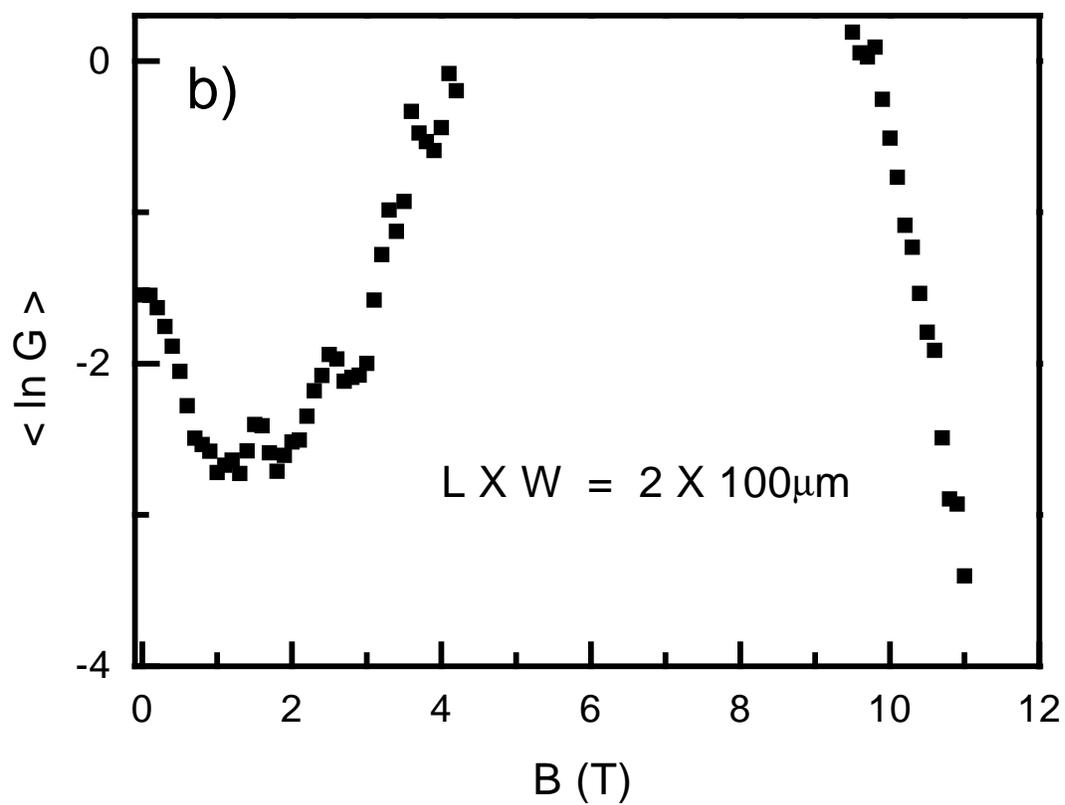
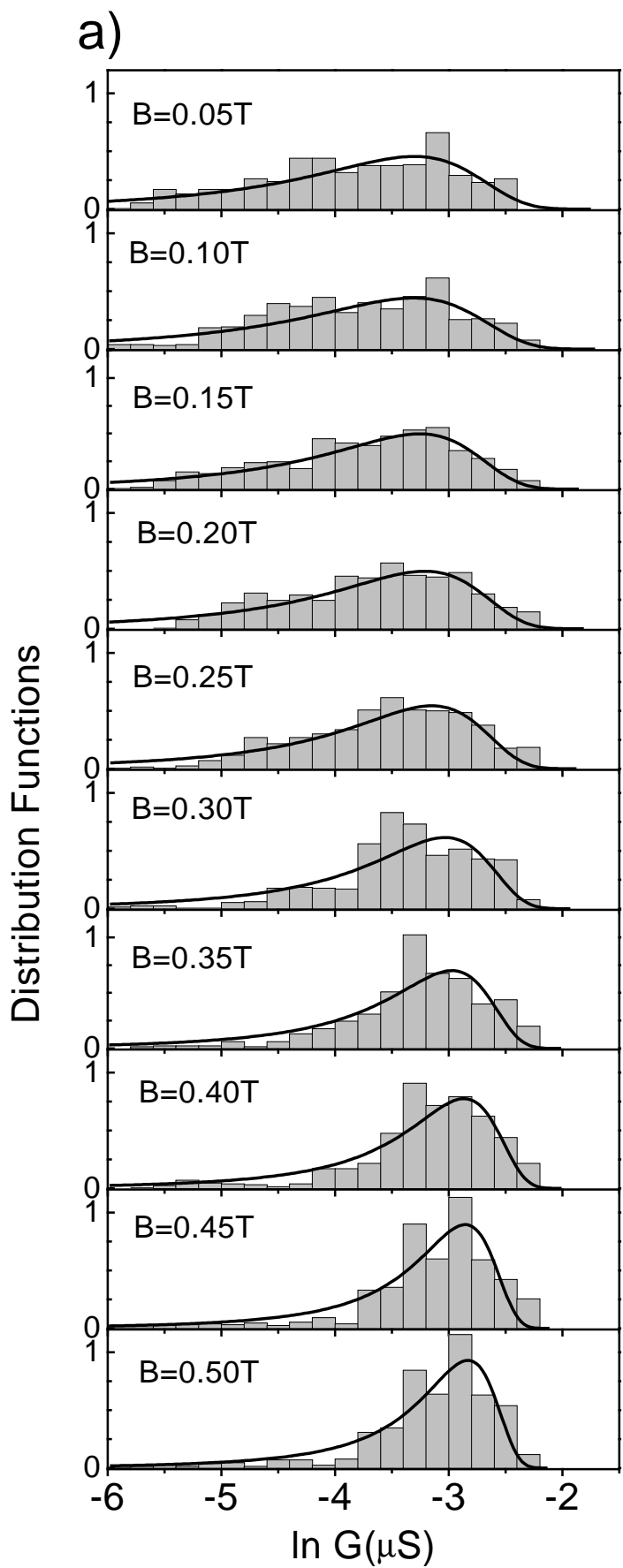
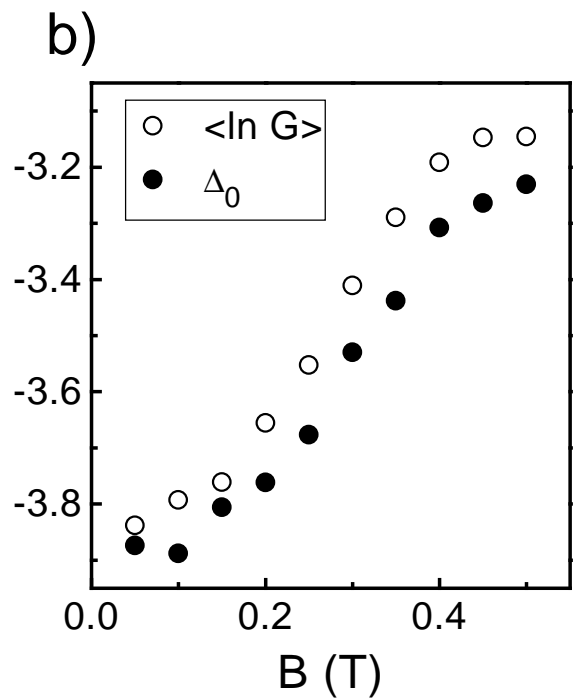
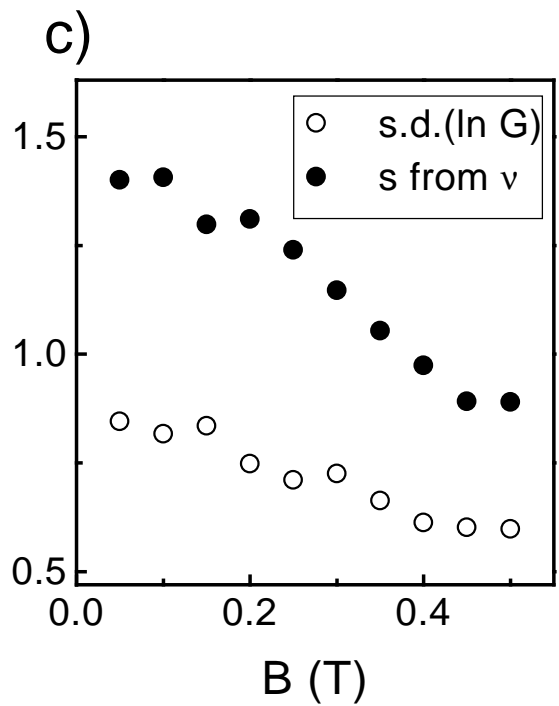
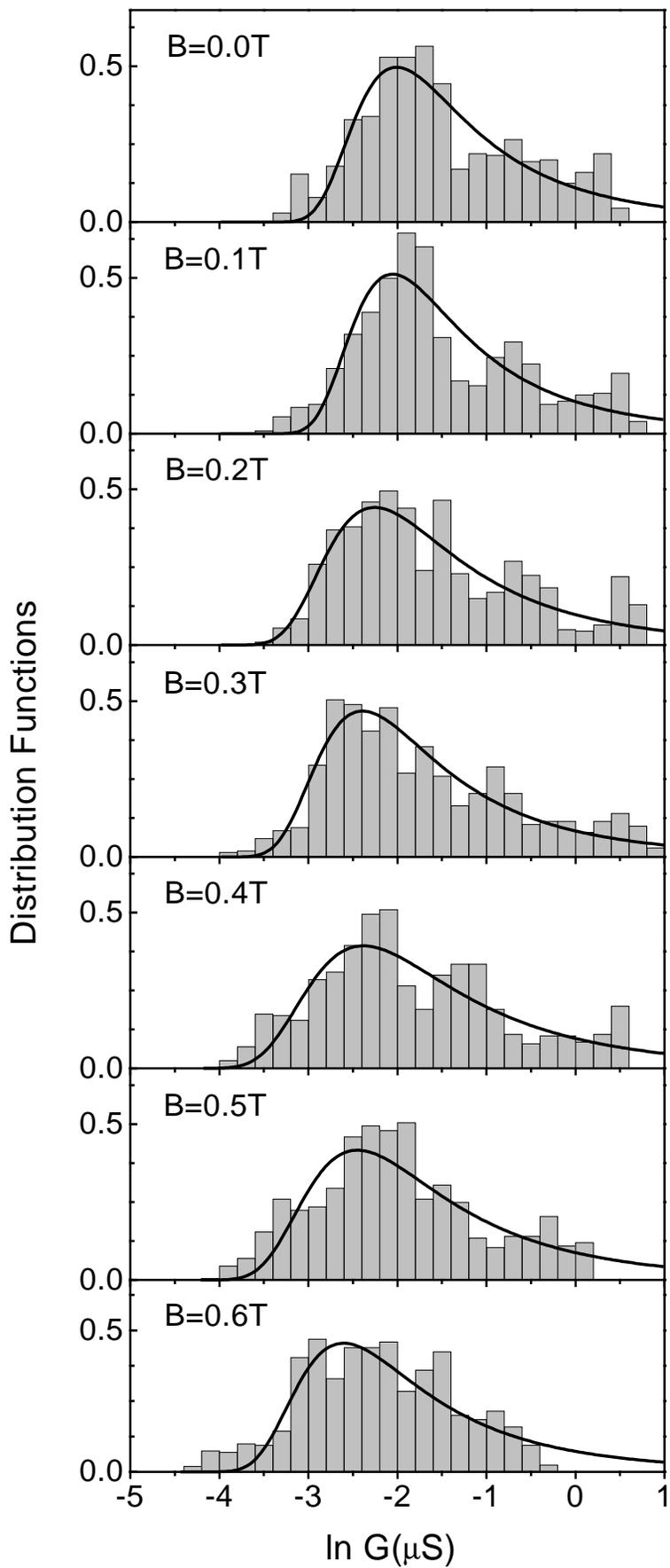
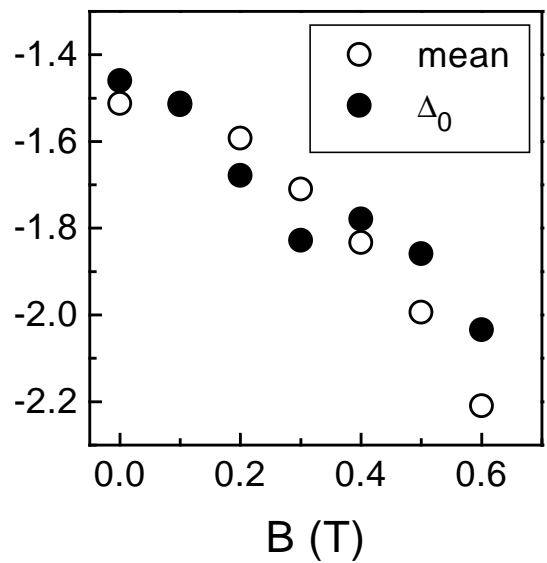
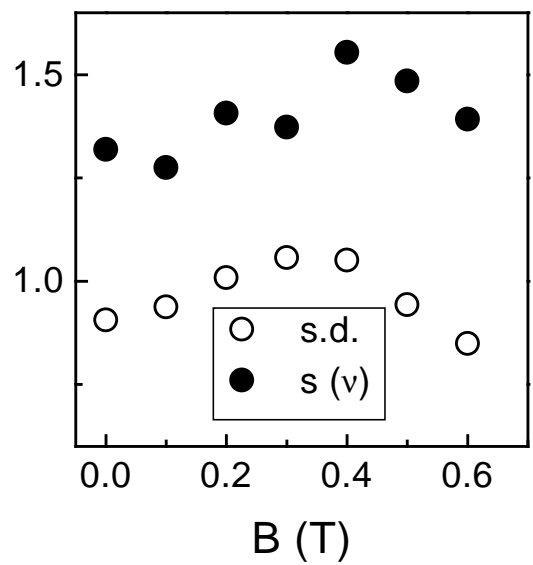
\end{document}